\documentstyle[aps,prb]{revtex}

\begin{document}

\title{Interacting electrons in polygonal quantum dots}

\author{C~E~Creffield$^1$, Wolfgang~H\"ausler$^2$, J~H~Jefferson$^3$ and
Sarben~Sarkar$^1$}

\address{
$^1$Dept. of Physics, King's College London, Strand, London, WC2R~2LS \\
$^2$Universit\"at Freiburg, Fakult\"at f\"ur Physik,
Hermann-Herder-Str. 3, D-79104 Freiburg, Germany \\
$^3$DERA, Electronics Sector, St. Andrews Road, Malvern,
Worcs. WR14~3PS }

\date{\today}

\maketitle

\begin{abstract}
The low-lying eigenstates of a system of two electrons confined within a
two-dimensional quantum dot with a hard polygonal boundary are obtained by
means of exact diagonalization. The transition from a weakly correlated
charge distribution for small dots to a strongly correlated "Wigner
molecule" for large dots is studied, and the behaviour at the crossover is
determined. In sufficiently large dots, a recently proposed mapping to an
effective charge-spin model is investigated, and is found to produce the
correct ordering of the energy levels and to give a good first approximation
to the size of the level spacings. We conclude that this approach is a
valuable method to obtain the low energy spectrum of few-electron quantum
dots.
\end{abstract}

\smallskip

\vspace*{0.25in} 

The success of modern nanoscale technology in the creation
and manipulation of quantum dots \cite{dots} has stimulated considerable
theoretical interest in elucidating the physical processes in these
structures. In the first analyses exact numerical diagonalizations were
carried out to study how many-body effects modify the energy level spectra 
\cite{bryant,merkt91,pfannkuche,peeters}, which may be measured by means of
non-linear transport spectroscopy \cite{weis}. As the number of electrons in
a quantum dot can be changed in a controlled way down to values of 
$N=1,2,3\dots$ \cite{dots,smalldots} they can be thought of as
``artificial'' H, He, Li etc. atoms, composed of interacting electrons
confined by an external potential. A vital difference between these
structures and real atoms, however, is the importance of correlations. Real
atoms can be described to good accuracy by independent electron models
within suitable mean field theories which can be systematically improved via
perturbation theory. This is not generally true for quantum dots for which
an independent electron approximation can give results which are not even
qualitatively correct. This behaviour is a consequence of the relatively
small kinetic energies in few electron quantum dots compared with their
mutual Coulomb interaction. For sufficiently low electron densities the
Wigner limit will be approached \cite{wigner,ceperley}, in which the
ground-state will adopt a quasi-crystalline form to minimize the interaction
energy. In non-translationally invariant systems this can occur at larger
`critical' densities than in the homogeneous 2D liquid \cite{tanatar}.
\newline

This feature of quantum dots containing low density electrons has been
exploited recently \cite{jeff} as a starting point complementary to
perturbative or mean field approaches that are reliable at weak interaction
strengths. As the electron density in these systems localizes around the
Wigner lattice points \cite{bedanov}, it is sensible to construct a
many-particle basis from antisymmetrized combinations of non-orthogonal
one-particle states localized at these sites. Orthogonalization leads to a
lattice model of strongly correlated electrons of Hubbard type, and the
subsequent elimination of the high-energy states of this model was then
shown to yield an effective ``$tJV$'' Hamiltonian. This effective
Hamiltonian has a very much smaller Hilbert space than that of the original
quantum dot enabling, in principle, the treatment of systems of many more
electrons than can be handled by direct diagonalization methods. The
ordering of the low energy levels as well as their relative separations
agree quite satisfactorily up to a scaling factor with data obtained by
exact diagonalization for a one-dimensional dot containing up to four
electrons \cite{jeff}. Here we examine the validity of the $tJV$--approach
in two-dimensional hard wall boxes. In the case of two (or higher)
dimensions, it should be noted that the number of energy minima in
configuration space, i.e. the number of peaks $N_{0}$ in the charge-density
distribution in the low density limit, can be larger than the number of
electrons. For example, for two electrons in a square box the ground-state
will consist of a superposition of states in which the electrons are located
on diagonally opposite vertices of the square, and so will exhibit a
four-peak structure. In this paper we consider three different box shapes:
triangular, square and hexagonal. For the case of two electrons the
eigenenergies and eigenfunctions are obtained by exact diagonalization, and
then compared with the predictions of the $tJV$ model.
\newline

We consider the Hamiltonian: 
\begin{equation}
H=\sum_{i=1}^{2} \ \left[ -\frac{\hbar ^{2}}{2m^{\ast }}\nabla _{i}^{2} +
V({\bf r}_{i})\right] \ + \ \frac{e^{2}}{4\pi \epsilon _{0}\epsilon _{r}}
\frac{1}{|{\bf r}_{1}-{\bf r}_{2}|}  
\label{ham}
\end{equation}
where we assume that the electrons can be described within the parabolic
band approximation by an effective mass $m^{\ast}$. The shape of the dot is
set by the confining potential $V({\bf r})$, which is zero within the dot
and infinite outside. The full wavefunction of the two-electron system may
be written as a product of a spinor and a spatial function: 
\begin{equation}
\Psi({\bf r}_{1},\sigma _{1};\ {\bf r}_{2},\sigma _{2})=
\psi({\bf r}_{1};{\bf r}_{2})\ \chi (\sigma _{1};\sigma _{2})
\end{equation}
where for a singlet (triplet) state $\psi$ is symmetric (antisymmetric)
under particle exchange. As the Hamiltonian (\ref{ham}) does not contain any
spin dependent terms it is only necessary to consider its action on the
spatial component of the wavefunction to obtain its eigenvalues. We choose
to use a basis of position eigenstates (a finite-difference method) as opposed
to the momentum eigenbasis used for example in \cite{bryant,kristian}. An
advantage of using this basis is that it is a simple matter to impose the
required homogeneous Dirichlet boundary conditions for dots regardless of
their shape, which can be very difficult to achieve in a momentum-space
basis. An exception to this is the case of the square shape, which we used
to check the reliability of the finite-difference approach by comparing with
results obtained using a momentum-space basis. The agreement was found to be
excellent.
\newline

Using the basis of position eigenstates amounts to replacing the spatial
continuum $({\bf r}_{1};{\bf r}_{2})$ with a four-dimensional mesh. The
wavefunction $\psi$ is then only evaluated at a discrete set of points: 
\begin{equation}
\psi (x_{1},y_{1};x_{2},y_{2})\rightarrow \psi _{i\ j\ k\ l}
\end{equation}
and the spatial derivatives in Eq.~(\ref{ham}) are replaced with symmetric
difference approximations: 
\begin{equation}
\frac{\partial^{2} \psi}{\partial x_{1}^{2}} \rightarrow 
\frac{\psi_{i+1 \ j\ k\ l} - 2 \ \psi _{i\ j\ k\ l} + 
\psi_{i-1\ j\ k\ l}}{(\Delta x)^{2}}, \qquad \mbox{etc.} 
\label{kinetic}
\end{equation}
The Hamiltonian can now be represented as a $N^{4}\times N^{4}$ matrix,
where $N$ is the number of mesh points per dimension. Since the Coulomb
interaction is diagonal in this representation, and the kinetic terms (\ref
{kinetic}) only operate between neighbouring mesh points, this matrix will
clearly be extremely sparse. This allows it to be stored in a very
economical fashion, and also permits an highly efficient use of Lanczos
diagonalization routines to obtain the lowest few eigenstates. As $N$ is
increased, and the separation between the mesh points decreases, the
spectrum of the discrete system will approach that of the continuum model
and the form of the mesh will be irrelevant. Finite-size effects can be
minimized, however, by choosing a mesh which matches the symmetries of the
potential as far as possible. For the case of a square dot we therefore
chose a square mesh, and a triangular mesh for the triangular and hexagonal
dots. On an Alpha workstation as many as $24^{4}$ lattice points could be
used, and in all cases we checked the convergence of the eigenvalues as the
number of lattice points was increased to this level.
\newline

In Fig.1 we present the ground state charge-distributions for the three
types of polygonal boundary for a selection of dot sizes $L$, where $L$ is
the side-length of the polygon. In each case the dot material was taken to
be GaAs, with an effective mass $m^{\ast }=0.067m_{e}$, and a relative
permittivity $\epsilon _{r}=10.9$, resulting in a Bohr radius of 
$a_{\rm B}=8.8$nm. For small dots the Coulomb interaction is weak on 
the scale of the kinetic energy, and the two-particle ground state resembles
the non-interacting ground state, with the charge distribution being peaked 
at the centre of the dot. Conversely, in very large dots the charge
distribution is strongly localized near the vertices of the dot where the
interaction is minimized, and the charge distribution practically vanishes
away from these maxima. In analogy to the formation of a Wigner crystal in
an infinite electron system, this configuration is referred to as a 
{\it Wigner molecule} \cite{kristian}. It is an important question to clarify
the characteristic scale $r_{\rm c}$ of the mean electron separation for which
the crossover between these two extreme cases takes place, since the actual
value of $r_{\rm c}$ cannot be obtained reliably using analytical
arguments. In 1D boxes $r_{\rm c}\approx 1.5a_{\rm B}$ has been found
from tracing the charge density distribution of the exact ground state 
\cite{kristian}. Using this technique we detect the change from the
non-interacting situation by observing the point at which the charge
distribution first shows a local minimum at the centre of the dot instead of
a maximum. In this way we estimate from Fig.1 that 
$r_{\rm c} \approx 10a_{\rm B}$ for all three dot geometries, which is almost
an order of magnitude larger than the values found in one dimension. This 
rough estimate is in reasonable agreement with the critical value of 
$r_{\rm c}\approx 35a_{\rm B}$ found by Ceperley for crystallization of the 
2D electron gas \cite{ceperley}. A somewhat larger value for $r_{\rm c}$ in 
two dimensions has been conjectured \cite{zpb} due to the considerably 
enhanced tendency for the electrons to surround one another. Although 
$r_{\rm c}$ marks the transition from the non-interacting regime, an 
additional qualitative change occurs at a somewhat longer length scale, at 
which the charge distribution develops $N_{0}$ sharply defined, well separated
maxima. It is to be expected that as the number of vertices is increased, and 
the polygon becomes closer to a circle, that this transition will occur at 
increasingly low densities. Accordingly we can observe this change at 
$\sim30a_{\rm B}$ for the triangle and square, but only at the larger value
of $\sim100a_{\rm B}$ for the hexagon. Before this transition occurs in 
the hexagonal dot the charge density shows a ring-like structure (as would
occur in a circular dot), and in this regime the spectrum indeed resembles 
that of a diatomic rotor composed of spin $1/2$ fermions, as can be seen in
Fig.6b. 
\newline

Fig.1 clearly shows how the electron localization takes place in
sufficiently large dots around specific sites in real-space.
Anticipating this behaviour motivated the mapping to a lattice model with a
combination of hopping, exchange, superexchange and Coulomb repulsion
processes. In Ref.\cite{jeff} a $tJV$ model: 
\begin{equation}
H^{tJV}={\cal P}\left[ \sum_{<i,j>,\sigma} -
t \ \left( c_{i\sigma }^{\dagger }c_{j\sigma }^{{}}+{\rm h.c.} \right) +
J \ \left( {\bf S}_{i}\cdot {\bf S}_{j}-\frac{n_{i}n_{j}}{4} \right) +
V \ n_{i}n_{j} \right] {\cal P} 
\label{h_tjv}
\end{equation}
was proposed to describe the low energy physics. Here ${\cal P}$ is a
projector eliminating doubly-occupied lattice sites, and $t$ and $J$ are the
standard hopping and Heisenberg terms between nearest neighbour sites. 
The $V$ term accounts for the nearest neighbor Coulomb repulsion and may be
written $V \approx e^{2}/4\pi \epsilon_{0}\epsilon_{r} L$ for large $L$.
Although first principles calculations of the energies $t$ and $J$ are
difficult, one can easily estimate the ordering of their magnitudes: 
\begin{equation}
|J| \ll |t| \ll V 
\label{esequence}
\end{equation}
For the three geometries considered here $H^{tJV}$ can be diagonalized
analytically to obtain the energy levels in terms of $t$, $J$ and $V$,
together with their corresponding spins. Let us first consider the case of
the square dot, which is described by two electrons moving on a four-site
lattice. Here the $V$ term in (\ref{h_tjv}) is important to discriminate
energetically between two electrons sitting on diagonally opposite vertices
as opposed to being on adjacent vertices. The lowest energy manifold of
states was derived in Ref.\cite{jeff} and, setting the ground-state energy
to zero, the manifold consists of a singlet ground state, two degenerate
triplet states at $(2\Delta +J)$, and a singlet at an energy of $4\Delta $
where $\Delta =2t^{2}/V$, and $\Delta \gg |J|$. The next set of levels are
separated from this lowest energy multiplet by an energy gap of the order of 
$V$. The transition from the almost constant density of states found in
small dots to a spectrum consisting of multiplets separated by large energy
gaps of this sort can be regarded as a signature for the crossover to the
Wigner regime \cite{hausler93}.
\newline

In Fig.2 we plot the lowest energy levels of the square dot as a function of
the dot size, normalized to the energy of the highest singlet state. The
overall structure of the spectrum agrees with that of the $tJV$ Hamiltonian 
(\ref{h_tjv}), and quantitatively reproduces the numerically exact spectra
obtained by Bryant for large dots \cite{bryant}. The doubly-degenerate
triplet state lies between the two singlet states and, asymptotically, the
spectrum becomes equidistant as $L\rightarrow \infty $. Remarkably, the
triplet levels are below $1/2$ for finite $L$ corresponding to a
ferromagnetic $J<0$. This is somewhat unexpected, since pair exchanges of
electrons would be expected to have antiferromagnetic couplings
\cite{herring62}. However, the analysis presented in \cite{jeff} does not 
{\it a priori} exclude negative values for the exchange coupling,
since the direct
exchange term might dominate the superexchange term. To exclude perturbative
influences on the ground manifold arising from higher excited states that
are ignored in the effective low energy description, we plot the decay of 
$|J/\Delta|$ versus dot size in Fig.3a. The excellent straight-line of the
data in this semilogarithmic plot clearly suggests that $J \propto \Delta \
e^{-L/r_{\rm c}}$, as opposed to the power law that perturbative
influences would give. The value of $r_{\rm c} \approx 52a_{\rm B}$, as
read off from Fig.3a, provides independent corroboration of the length scale
characterizing the transition to the Wigner state estimated earlier from the
charge density distributions.
\newline

The dominant energy scale $\Delta$ can be roughly estimated using the
pocket state picture \cite{zpb} together with the WKB approximation: 
\begin{equation}
\Delta \sim {\rm e}^{-S_{0}} 
\label{wkb}
\end{equation}
where the classical Euclidian action:
\begin{equation}
S_{0} = \int_1^2{\rm d}\vec{q} \; \sqrt{2mn(v(\vec{q})-v(\vec{0}))}
\end{equation}
is taken along the path 
$\vec{q}:({\bf r}_1^{(1)},{\bf r}_2^{(1)}) 
\to({\bf r}_1^{(2)},{\bf r}_2^{(2)})$
that carries $n$ electrons 
between two adjacent energy minimum configurations `(1)' and `(2)' 
($n=2$ for the example of the square and $n=1$ for the 
triangle). Here, $v(\vec{q})= e^2/4\pi\epsilon_{0}\epsilon_{r}
|{\bf r}_{1}-{\bf r}_{2}|$. For the square the path corresponds to a 
$\pi/2$ rotation of the two electrons around the centre of the square, 
yielding:
\begin{equation}
S_{0}^{\mbox{\tiny square}}\approx 0.79\sqrt{r_{\rm s}}\quad.
\label{wkb_sqr}
\end{equation}
The parameter $r_{\rm s}$ is the typical scale of electron separation 
(which for two electrons is equal to the dot size $L$).
In Fig.3b we show $\Delta$
versus $\sqrt{L}$ on a semilogarithmic plot and verify that $\Delta \sim 
{\rm e}^{-\sqrt{L/\xi }}$. From the slope we extract a value for
$\xi=1.64a_{\rm B}$ that is in excellent agreement with the pocket state
prediction of $\xi =1.60a_{\rm B}$ from Eq.\ref{wkb_sqr}.
\newline

The triangular dot is described by two electrons moving on a three-site
lattice, and hence the $V$ term is irrelevant as it will just give rise to
an overall shift in energy levels. The resulting $tJ$ model has a more
complicated ground-state manifold than for the square dot, and is shown in
Fig.4a. This is also exactly the ground-state manifold for the hexagonal
dot, where the $V$ term is again of importance. We can again employ the
pocket state WKB theory to estimate the magnitude of the dominant energy
scale $t$. For the triangle we have: 
\begin{equation}
S_0^{\mbox{\tiny triangle}}
\approx 0.42\sqrt{r_{\rm s}} 
\label{wkb_tri}
\end{equation}
for one electron hopping along the edge to the empty site. In Fig.5 we
present a semilogarithmic plot of $t$ versus $\sqrt{L}$ for the triangle
which again confirms that the scaling of the dominant energy is of the form 
$t \sim {\rm e}^{-\sqrt{L/\xi}}$. The value of $\xi=4.45 a_{\rm B}$
measured from this plot compares reasonably with $\xi=5.67 a_{\rm B}$ as
predicted by WKB theory from Eq.\ref{wkb_tri}.
\newline

In Fig.6 we present the energy levels obtained by the diagonalizations
(again scaled by the energy of the highest singlet state) for the triangle
and the hexagon. The sequence and the asymptotic ratios of the energy
separations again agree with that predicted by the $tJV$ model. In contrast
to the earlier result, however, it is not possible to fit these results with
a single $J$. For example, for both geometries the higher triplet approaches
its asymptotic value 4/3 from above, implying $J>0$, whereas the lower
triplet approaches 1/3 from below, requiring a ferromagnetic $J$.
This behaviour indicates that although the $tJV$ Hamiltonian is able to
adequately predict the gross features of the low energy spectrum, it does
not, in fact, provide a complete description of the dynamics occurring in
the triangular and hexagonal dots. 
As this Hamiltonian was obtained as a reduced version of a more general
model of Hubbard type \cite{jeff} this raises the possibility that during
the reduction procedure some terms were dropped which may be of importance
in these situations. To investigate this further we again consider the
lattice model with three sites forming an equilateral triangle and occupied
by two electrons. With one orthogonal state per site the Hamiltonian is
a generalised Hubbard model: 
\begin{equation}
H=\sum_{ij\sigma }t_{ij}c_{i\sigma }^{\dagger }c_{j\sigma } +
\frac{1}{2}\sum_{ijkl \sigma \sigma^{\prime}} \ U_{ijkl}
c_{i\sigma }^{\dagger} c_{j\sigma^{\prime }}^{\dagger}
c_{l\sigma^{\prime }} c_{k\sigma} 
\label{lattice}
\end{equation}
where $t_{ij}$ are the one electron matrix elements coming from the kinetic
energy and the confining potential and $U_{ijkl}$ are Coulomb matrix
elements. We now consider {\em all} contributions to the low-lying manifold
for which there is no double occupation of sites. (These are much higher in
energy and their effect may be accounted for by second-order perturbation
theory where they give rise to an antiferromagnetic superexchange term, as
described in Ref.\cite{jeff}). The largest intersite Coulomb term has matrix
elements $U_{ijij}\equiv V$ and the corresponding (ferromagnetic) exchange
term has matrix elements $U_{ijji} \equiv J_{\rm F}/2$. These terms are
already included in the $tJV$ Hamiltonian. All other terms in 
(\ref{lattice}) involve hopping of electrons and fall into the following 
three classes:

\begin{enumerate}
\item  $(i=j=l:k\rightarrow j)$
\[
\frac{1}{2}\sum_{ij \sigma} U_{iiji} c_{i \sigma}^{\dagger} 
c_{i\bar{\sigma}}^{\dagger} c_{j\bar{\sigma}} c_{i\sigma} = 
\frac{1}{2} \sum_{ij \sigma} U_{iiji} n_{i\sigma }
c_{i\bar{\sigma}}^{\dagger} c_{j\bar{\sigma}}
\hspace{1in}
\]
This is a spin-dependent hopping term which {\em always} takes us to a high
energy state since it involves double occupation (e.g. electron hops from
site 1 to site 2 provided site 2 is occupied with an electron of opposite
spin). It thus contributes to superexchange by lowering the energy of
singlets. This term, which renormalizes $J$, is also 
discussed Ref.\cite{jeff}.

\item  $(i=k:j\neq l\neq i)$ 
\[
\frac{1}{2} \sum_{ijl\sigma \sigma ^{\prime}} U_{iijl}
c_{i\sigma}^{\dagger} c_{j\sigma ^{\prime}}^{\dagger }
c_{l\sigma^{\prime}} c_{i\sigma}=
\frac{1}{2}\sum_{ijl\sigma } U_{iijl} n_{i} c_{j\sigma}^{\dagger}
c_{l\sigma}
\]
This is potentially an important term since it operates in the ground
manifold. However, for two electrons in a triangle, we can set $n_{i}=1$ (ie
site $i$ is always occupied for states in which 
$c_{j\sigma}^{\dagger} c_{l\sigma}$ does not give zero) and hence 
this term merely
renormalizes the kinetic energy ($t$)-term.

\item  $(i=l:j\neq k\neq i)$
\begin{eqnarray*}
\frac{1}{2}\sum_{ijk\sigma \sigma^{\prime}} U_{ijki}
c_{i\sigma}^{\dagger} c_{j\sigma^{\prime}}^{\dagger}
c_{i\sigma^{\prime}} c_{k\sigma}
&=&-\frac{1}{2} \sum_{ijk \sigma \sigma^{\prime}} U_{ijki}
c_{i\sigma}^{\dagger} c_{i\sigma^{\prime}} 
c_{j\sigma^{\prime}}^{\dagger} c_{k\sigma} \\
&=&-K\sum_{ijk\sigma \sigma^{\prime}} 
c_{i\sigma}^{\dagger} c_{i\sigma^{\prime}} 
c_{j\sigma^{\prime}}^{\dagger} c_{k\sigma}
\end{eqnarray*}
where the last step follows since all $U$'s are the same ($K=U/2$).
\end{enumerate}

This last term is potentially important since it operates in the ground
manifold and it cannot be accounted for simply as a renormalization of the
other parameters (i.e. $J$ and $t$). It is also spin dependent, behaving
differently when operating on singlet states from triplet states. Consider a
base state in which there is a spin up electron on site $i$, a spin down
electron on site $k$ and no electron on site $j$. This operator will then
move an electron from site $k$ to site $j$ and then flip the spins of sites 
$i$ and $j$. It is thus a combined hop and spin-flip, as shown in Fig.7a. On
the other hand, when it operates on a state where both electrons have the
same spin it only performs the hop. 
Furthermore, an estimate of the
magnitude of $K$ shows that it is comparable with $J$.
\newline

Retaining all terms gives the effective model: 
\begin{eqnarray}
H^{\rm eff} &=& H^{tJ}-K \sum_{ijk\ \sigma \sigma^{\prime}}
c_{i\sigma}^{\dagger} c_{i\sigma^{\prime}} 
c_{j\sigma^{\prime}}^{\dagger} c_{k\sigma } 
\label{tJVK} \\
&=& H^{tJ}- 2K \sum_{ijk\ \sigma } \left( 
{\bf S}_{i} \cdot {\bf S}_{j}+
\frac{1}{4} \right) c_{j \sigma}^{\dagger} c_{k\sigma}
\end{eqnarray}
where $K$ is a positive coupling and the $V$-dependence has been dropped
since it gives the same energy contribution for all states. Introducing this
additional term alters the eigenenergies as shown in Fig.4b, and this
alteration is of exactly the correct form to account for the energy level
separations obtained from the numerical diagonalizations. For both dot
geometries, fitting the results with these parameters yields a positive
value for $K$ as expected, and a positive (antiferromagnetic) $J$ of similar
magnitude. 
\newline

We can also show that a $K$-term is also important for the low-lying
manifold of other polygonal quantum dots. Consider, for the example, the
regular hexagonal geometry for which the exact numerical low-lying spectrum
was described earlier. Starting with an extended Hubbard model and taking the
`atomic' limit gives a $24$-fold degenerate ground manifold of states,
corresponding to six equivalent positions in which the two electrons are
directly opposite each other and a factor four for spin. In second-order the
degeneracy is partly lifted giving an effective Hamiltonian with `ring'
terms, which correspond to a simultaneous rotation of the electrons by
$\pm \pi /3$ about the centre of the dot. In fourth-order we get the usual
superexchange ($J$) term but, in addition, we get a $K$-term which also
corresponds to a simultaneous rotation of $\pm \pi /3$ but now also involves
a spin-flip, as shown in Fig.7b. The final effective Hamiltonian has the
form:
\begin{equation}
H^{\rm eff}=\sum_{<i,j>} \left[ -\Delta 
\left( R_{\frac{\pi}{3}} + R_{-\frac{\pi}{3}}\right) +
J \left( {\bf S}_{i} \cdot {\bf S}_{j} - \frac{1}{4} \right)
-K \left( {\bf S}_{i}\cdot {\bf S}_{j} + \frac{1}{4}\right) 
\left( R_{\frac{\pi}{3}} + R_{-\frac{\pi }{3}}\right) \right] n_{i}n_{j}
\label{DJK}
\end{equation}
where the summation is over all six pairs of opposite sites. The effective
Hamiltonian (\ref{DJK}) describing the low-lying states of the hexagon is
isomorphic to the effective Hamiltonian for the triangular dot 
(Eq.\ref{tJVK}) with $\Delta $ playing the role of $t$.
This may be seen explicitly
by writing down expressions for the effective Hamiltonian matrices in their
respective localized bases. Thus the low-lying eigenstates of the triangular
and hexagonal dots are in one-to-one correspondence, in agreement with the
numerical results.
\newline

Note that by a similar reasoning we can generate effective $K$-terms for any
polygonal dot. With the exception of the square dot, such terms are
necessary in order to give quantitative agreement with the exact low-lying
spectrum. This is not the case for the square since, as shown in Fig.7c, a
rotation of $\pi /2$ followed by a spin-flip is equivalent to a rotation of 
$-\pi /2$ with no spin-flip. The effect is, therefore, to merely renormalize
the second-order ring processes.
\newline

In conclusion we have examined the behaviour of the lowest energy levels of
two electrons confined to two-dimensional polygonal quantum dots. For
sufficiently large dots the ground-state charge distribution shows a
quasi-crystalline structure, which can be used as the basis for mapping the
system to an effective $tJV$ lattice model. This model is found to give the
correct ordering of energy levels and to give a good first approximation to
the energy level spacings. The dominant energy scale of the systems can be
estimated semi-classically to good accuracy, and although it is hard to
obtain estimates for the remaining parameters, they decrease exponentially
with the size of the dot and so this description becomes increasingly
reliable in large dots. For the case of triangular and hexagonal dots,
however, it was evident that it was necessary to retain the three-site terms
normally neglected in the derivation of the effective model to account for
the detailed behaviour of the lowest energy levels. The lattice description
of the dot considerably simplifies the calculation of the energy spectrum,
and provides an appealing interpretation of the low-energy excitations
occurring in these structures. It is still desirable to investigate
different electron numbers and dot geometries to check the universality of
our results concerning the critical density characterizing the cross-over
into the Wigner regime. 
\vspace{0.25in} 

{\bf Acknowledgements~:} CEC is grateful for financial support from the
Leverhulme Foundation and WH acknowledges support by the Deutsche
Forschungsgemeinschaft through SFB 276. EU support from the TMR programme is
also gratefully acknowledged.

\vspace{0.25 in} \noindent{\bf Figure captions} \newline
\noindent Fig.1: Ground state charge distributions for the three types of
quantum dot. Dot sizes are: (a)50 nm (b)100 nm (c)800 nm \newline
\noindent Fig.2: Lowest energy levels for a square dot. \newline
\noindent Fig.3: (a)~Decay of $J/\Delta$ with dot size $L$; (b)~Decay of 
$\Delta$ with $L$, line of best-fit is shown. \newline
\noindent Fig.4: (a)~Energy level structure obtained by solution of the $tJ$
model. Level degeneracies are shown in brackets. (b)~Modifications to the
energy levels produced by the addition of the $K$ term (see text). \newline
\noindent Fig.5: Decay of $t$ with $L$ for a triangular dot, line of
best-fit is shown. \newline
\noindent Fig.6: Energy level structures: $\bullet$ denotes singlet states, 
$\circ$ denotes triplet states. Level degeneracies are shown in brackets.
(a)~Triangular dot: the asymptotic decay to the values predicted by the $tJ$
model is very evident. (b)~Hexagonal dot: the decay to the symptotic values
is clearly less rapid than for the triangle. The arrows mark the energy
levels of a rigid rotor system. \newline
\noindent Fig.7: K--processes. (a)~Triangular dot: hop followed by spin-flip.
(b)~Equivalent process for hexagonal dot: $\pi/3$ rotation followed
by spin-flip. (c)~Square dot: $\pi/2$ rotation followed by spin-flip,
equivalent to $-\pi/2$ rotation without spin-flip.

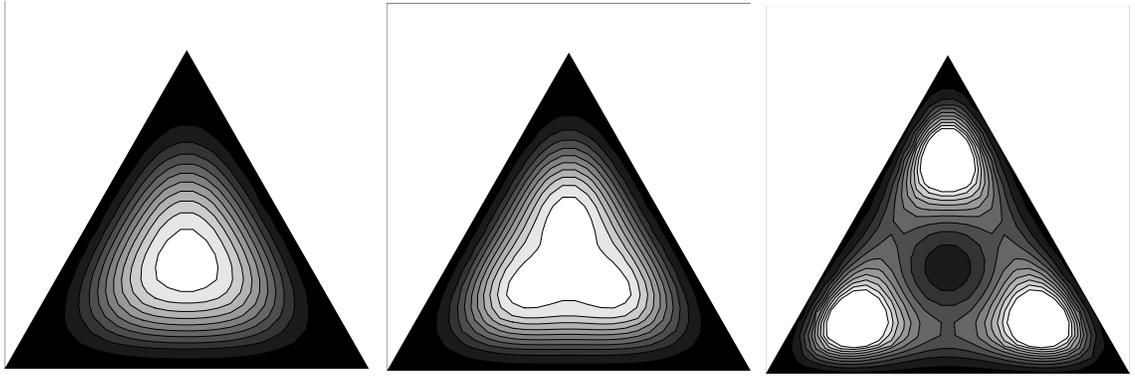
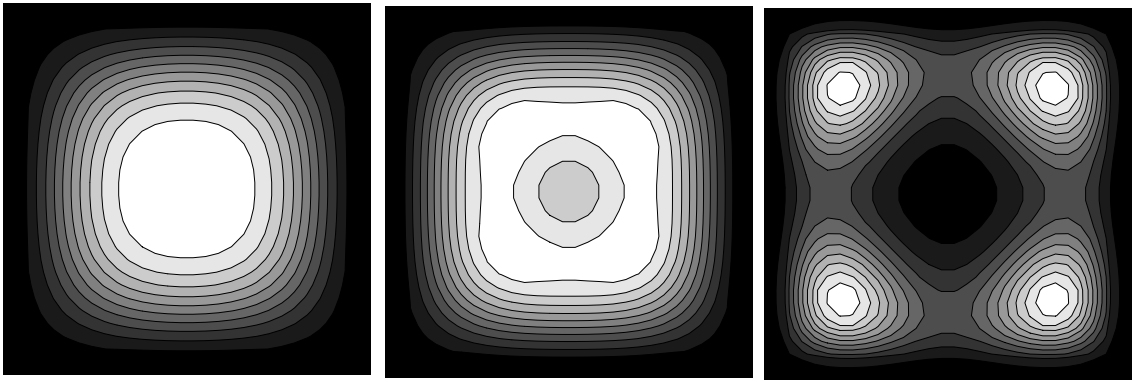
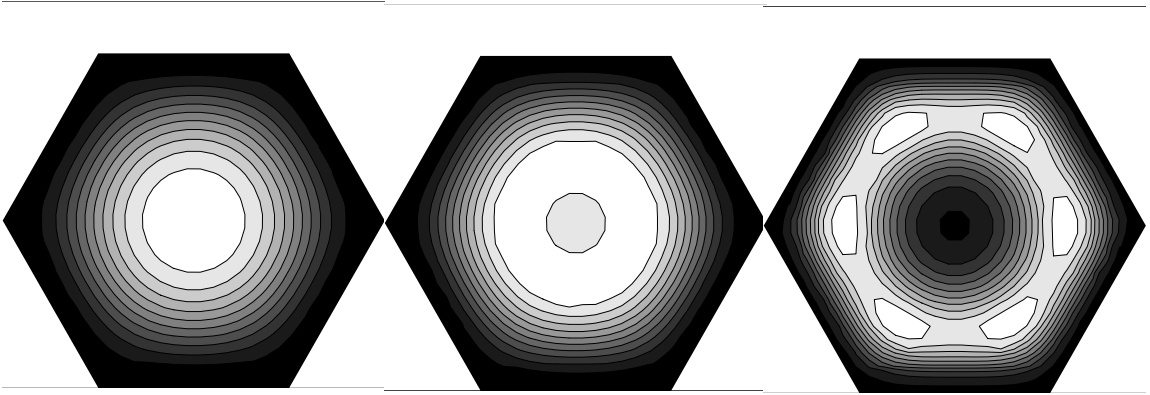
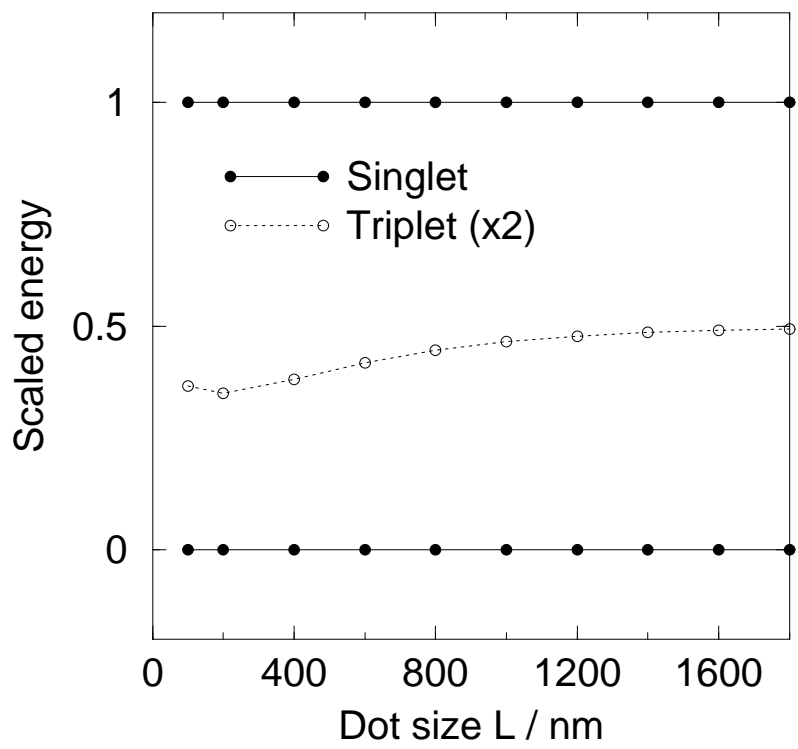
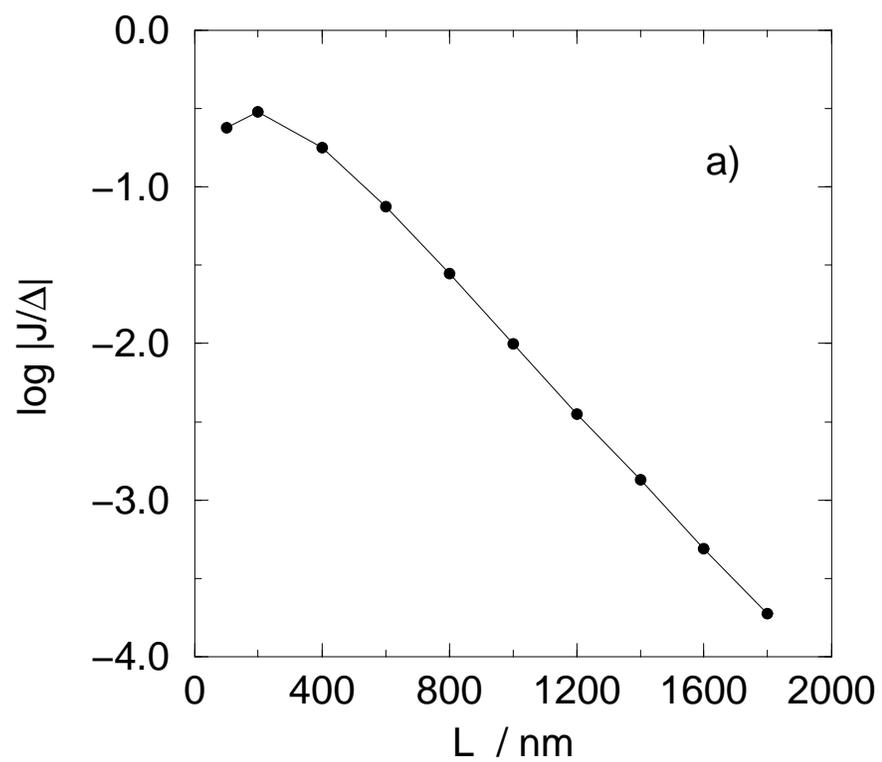
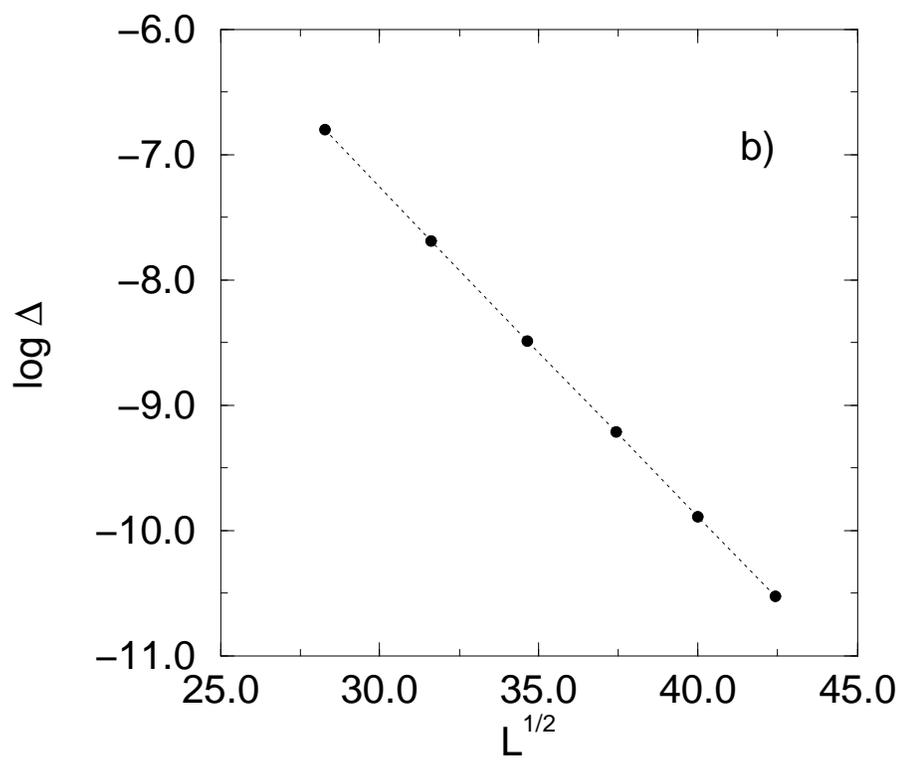
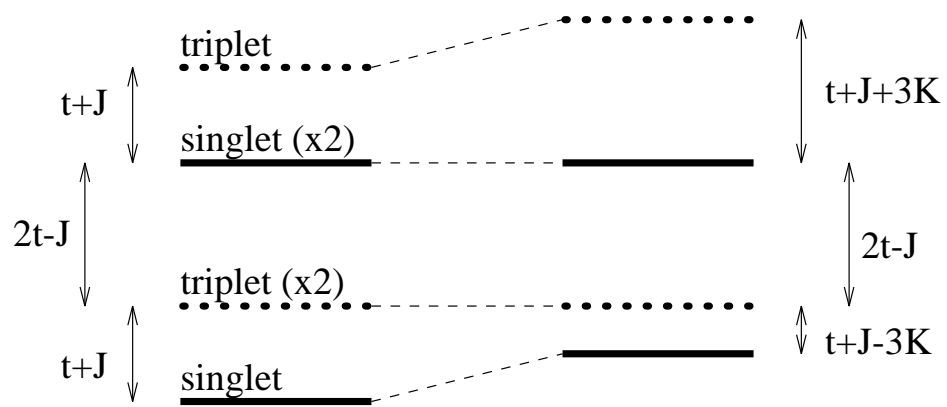
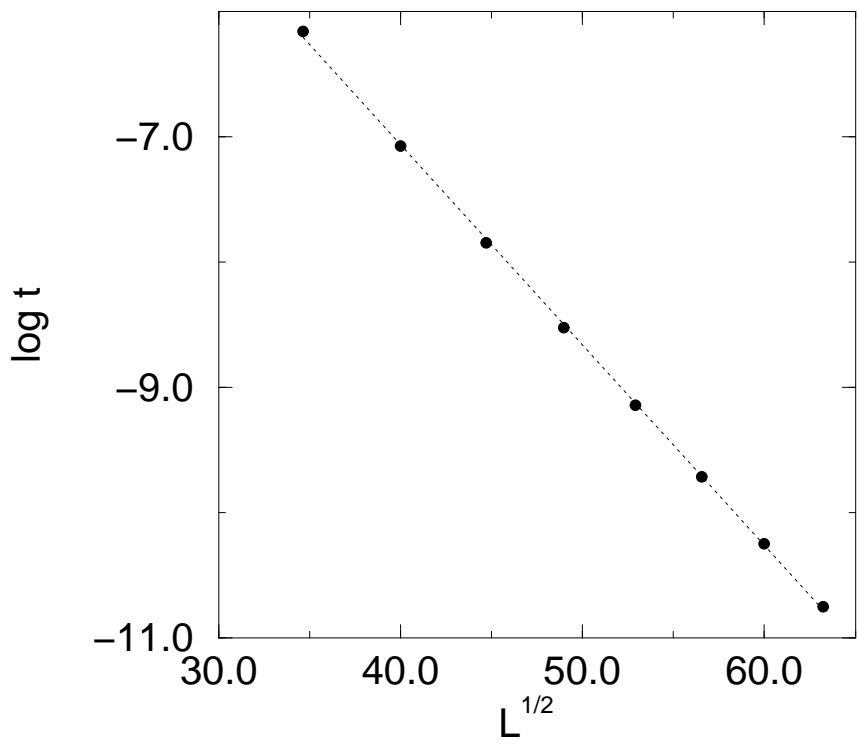
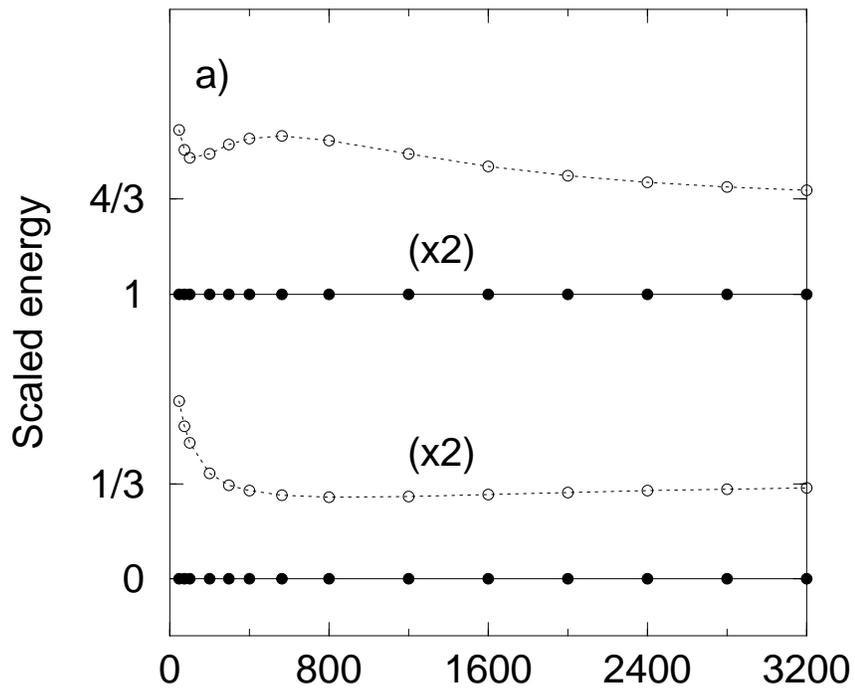
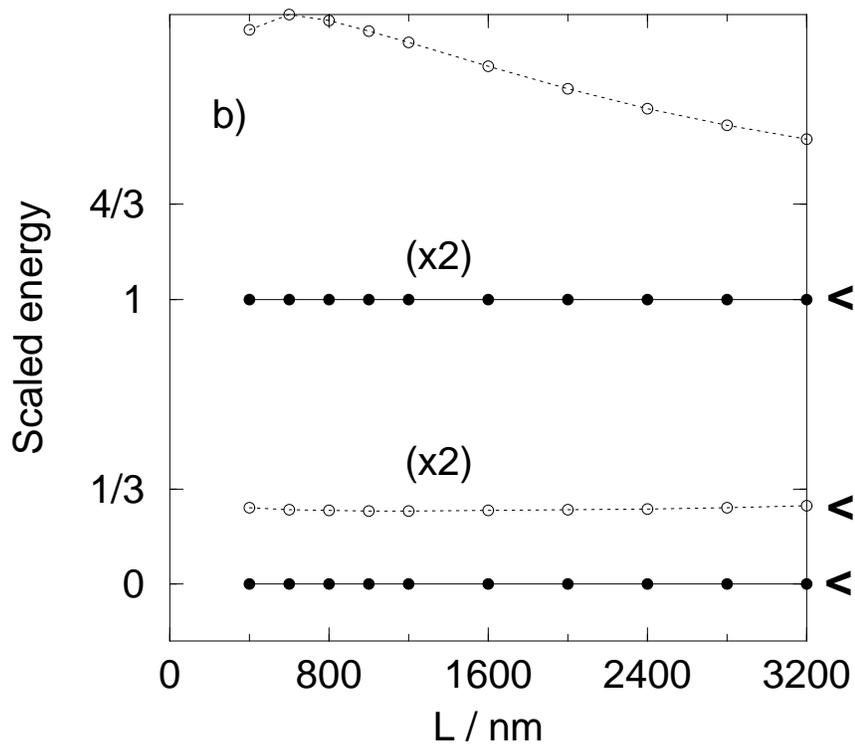
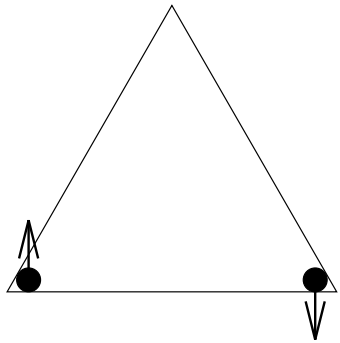
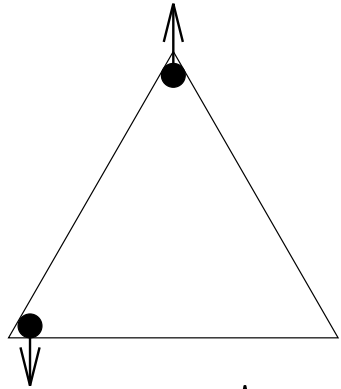
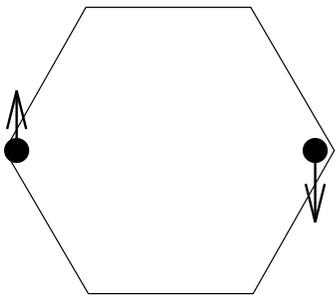
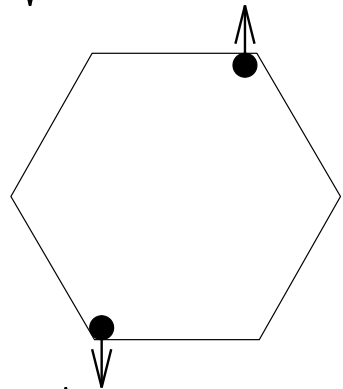
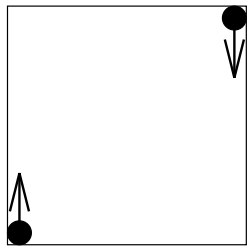
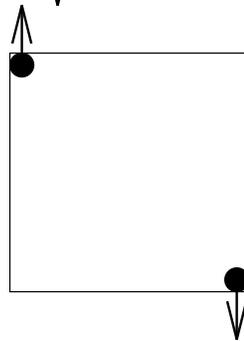
\end{document}